\documentclass[preprint,prb,twocolumn,
10pt
]{revtex4}
\usepackage{amsmath}
\usepackage{amssymb}
\usepackage{amsthm}
\newcommand{{\bfa}}{\mbox{\boldmath$a$\unboldmath}}
\newcommand{{\bfmu}}{\mbox{\boldmath$\mu$\unboldmath}}

\begin{document}

\title{\Large Initial accelerations of pulsars caused by external kicks} 
\author{Ricardo Heras}
\email{ricardoherasosorno@gmail.com}
\affiliation{Preparatoria Abierta de la SEP en Toluca Edo.~de M\'exico}
\begin{abstract}
\noindent 
Using energy conservation we show that if a sudden external force of unknown nature causes a newborn pulsar of mass $M\!=\!1.4 M_\odot$ and  radius $R\!=\!10$ km to have the acceleration $a_0$ then we can derive the relation $a_0\!\simeq \!k^{-1}vB_0^{-2/3},$ where $v$ is the velocity of the pulsar, $B_0$ its initial magnetic field and $k\!\approx\!10^{-17}$G$^{-2/3}$s. This relation predicts 
that a newborn pulsar with $v\!=\!450$ km experienced the acceleration $a_0\!\simeq10^{13}$g when $B_0\!\simeq\!10^{13}$G. An external force producing such an acceleration seems not to be physically feasible. This pulsar could have experienced the more realistic acceleration $a_0\!\simeq\!10^{9}$g if 
$B_0\!\simeq\!10^{18}$G. But this huge magnetic field seems to be unrealistic. 
\end{abstract}
\maketitle
The extremely violent explosion of a supernova is triggered by the gravitational collapse of an evolved degenerated core of a massive star $(\gtrsim  10 M_\odot)$ and trows a fast pulsar as a remanent [1]. The velocity of pulsars have a mean value of 450 km/s [2,3]. The pulsar velocities are, on average, one order of magnitude greater than those of their progenitor stars. No completely satisfactory explanation for the origin of the 
high velocity of pulsars has been proposed yet [4]. 

In this note we explore the possibility that the conversion of the radiative energy (originated by the action of an external force) into kinetic energy 
could explain the observed high velocity of pulsars. The basic idea is the following. According to Maxwell's theory, if a sudden external force of unknown nature causes a pulsar of mass $M$ and magnetic field $B_0$ to have an acceleration of magnitude $a_0$ for a short interval of time $\tau$ then the pulsar will radiate. The associated radiative energy  is converted into kinetic energy and then the pulsar is expected to acquire its high velocity. 

The power radiated by a suddenly accelerated pulsar of radius $R$ and a magnetic field $B_0$ can be estimated by   
\begin{equation}
P\!\simeq \!\frac{\!B_0^2R^6 a_0^2}{15c^5 \tau^2},
\end{equation}
which can be obtained from the Larmor formula for a suddenly accelerated magnetic dipole moment [5,6]:
\begin{equation}
P=\frac{4\mu_0^2\dot a^2}{15c^5},
\end{equation}
by assuming the estimate $\dot a\sim a_0/\tau$ and making the substitution $\mu_0\!=\!B_0R^3/2$.
Equation~(2) assumes that the magnetic dipole moment $\bfmu$ is perpendicular to the time derivative of the acceleration $\dot {\bfa} $. For the case in which $\dot {\bfa} $ and $\bfmu$ are parallel we have $P=2\mu_0^2\dot a^2/(15c^5).$ Equation~(1) assumes a nonrelativistic motion for the magnetic dipole. Thus the electromagnetic energy radiated by the nascent pulsar during the time $\tau$ can be estimated by  
\begin{equation}
{\cal E}_{\rm rad}\!\simeq \frac{\!B_0^2R^6 a_0^2}{15c^5 \tau}.
\end{equation}
If this radiative energy is transformed into kinetic energy of the pulsar, ${\cal E}_{\rm k}\sim M v^2,$ then 
\begin{equation}
\frac{\!B_0^2R^6 a_0^2}{15c^5 \tau}\simeq M v^2,
\end{equation}
which implies 
\begin{equation}
v^2\simeq a_0^2\tau^2\bigg(\frac{B_0^2R^6}{15Mc^5 \tau^3}\bigg).
\end{equation}
If the time $\tau$ is expressed as 
\begin{equation}
\tau=\bigg(\frac{B_0^2R^6}{15Mc^5}\bigg)^{1/3},
\end{equation}
then Eq.~(5) becomes $v^2\simeq a_0^2\tau^2$ which implies the following expression for the pulsar velocity
\begin{equation}
v\simeq a_0\tau.
\end{equation}
Alternatively, if we assume Eq.~(7) and substitute it into Eq.~(4) then we obtain Eq.~(6). Clearly, when the external force is applied 
during times $T\!>\!>\!\tau$ then the radiative effects implied by Eq.~(4) would be unimportant.
A formal treatment for the dynamics of a  suddenly accelerated magnetic dipole including radiation reaction predicts the kick velocity 
$v\simeq a\tau$ [7] where $a=F/m$ with  $m$ being the dipole mass and $F$ the magnitude of a constant force acting on the dipole during the time $\tau$ (expressed in terms of the magnetic moment). According to Eq.~(7), a nascent pulsar of mass $M= 1.4 M_\odot$ and with an initial magnetic field $B_0=10^{13}$G has associated the characteristic time 
\begin{equation}
\tau=4.6\times 10^{-9}{\rm s}. 
\end{equation}
This is the time taken for light to travel 2.1 m! We can also estimate the magnitude of the sudden acceleration acting on the nascent pulsar. For a  pulsar with velocity  $v=450$ km/s and characteristic time $\tau=4.6\times 10^{-9}$s, Eq.~(7) implies the initial acceleration 
\begin{equation}
a\simeq 10^{13}{\rm g}.
\end{equation}
Equation (7) can alternatively be written as 
\begin{equation}
v\simeq k a_0B_0^{2/3},
\end{equation}
where $k\!=\![R^6/(15mc^5)]^{1/3}$. If $R\!=\!10$ km and $M\!= \!1.4 M_\odot$ then $k\!\approx\!10^{-17}{\rm G}^{-2/3}$s. As may be seen, the pulsar velocity in Eq.~(10) depends on the initial acceleration and on the initial magnetic field. From Eq.~(10) we obtain 
\begin{equation}
a_0\simeq \frac{vB_0^{-2/3}}{k}.
\end{equation}
For pulsars of characteristic ages $\leq 10^{6}$\:yr we can assume that the values of the initial magnetic field $B_0$ can be approximated by their current values on its surface. On the other hand, the space velocity $v$ can roughly be estimated by its associated transverse velocity. In the first version of this note we presented a table including the initial accelerations of a sample of 99 pulsars with ages $\leq 10^{6}$\:yr. 
The mean value of the transverse velocities of the sample was 337 km/s and the corresponding mean value of the magnetic fields was $5\times10^{12}$G. The mean value of the initial acceleration was found to be $3\times10^{13}$g.

Nevertheless, the mechanism suggested here to explain the pulsar velocities can strongly be questioned because of the extremely small values for $\tau$.  From physical considerations the speed of light $c$ would be the fastest speed to administer the overall kick to the pulsar, whatever may be the cause of the kick. The time $t_c\!\approx\! 3.4\!\times\!10^{-5}$s is the time taken for light to cross the newborn pulsar. One should conclude that the electromagnetic radiative effects described by Eq.~(4) are unimportant because in this case $t_c\!>\!>\!\tau$.  Moreover, external forces producing accelerations $a_0\!\simeq\!10^{13}$g in characteristic times $\tau\!\simeq\!10^{-9}$s  seem not to be physically feasible.

The radiative mechanism proposed here could be physically acceptable if the characteristic time in Eq.~(6)
is at least of the order of the time taken for light to cross the pulsar: $\tau\!\sim\!\! 10^{-5}$s. From Eq.~(6) we can see that this value for $\tau$ could be obtained by assuming a huge initial magnetic field $B_0\!\sim\!10^{18}$G. With this huge initial field Eq.~(11) gives the initial acceleration $4.6\times 10^9$g for the velocity 450 km/s. The required external kick would produce an acceleration four orders of magnitude less than that given in Eq.~(9)  and its characteristic time would be four orders of magnitude more than that given in Eq.~(8).

From theoretical considerations, the physical upper limit to the pulsar magnetic field follows from the virial theorem of magnetohydrostatic
equilibrium [8,9]. The magnetic energy of the pulsar: ${\cal E}_m\!=\![4\pi R^3/3]B_0^2/(8\pi),$ can never exceed its gravitational binding energy: ${\cal E}_g\!\sim\! GM^2/R,$ that is, ${\cal E}_m\!\lesssim\!{\cal E}_g$ which implies magnetic fields in the bulk of pulsars satisfying  $
B_0\!\lesssim\!\sqrt{6G}M/R^2.$ Substituting $G\!=\!6.67\!\times \!10^{-8}$dyne cm$^2$/gr$^2$, $M\!=\! 2.8\! \times \!10^{33}$gr and $R\!=\!10^6$cm we obtain $B_0\!\lesssim \!10^{18}\;{\rm G}.$ In connection with an initial magnetic field of the order of $10^{18}$G, Lerche and Schramm [10] have pointed out that:  ``...a field of this order will rapidly decay because optical photons produce pairs or ambient electrons give rice both to magnetic bremsstrahlung and to synchrotron radiation$-$both of which deplete the field."

However, the idea that pulsars had extreme initial magnetic fields of the order of $10^{18}$G seems to be unrealistic and involves serious difficulties. For example, it is not clear how millisecond pulsars with current magnetic fields of the order of $10^{8}$G had initial magnetic fields of the order of $10^{18}$G.

\vskip 5pt
The author is grateful to Jos\'e A. Heras who suggested the basic idea developed in this paper. Useful criticisms received from Prof. Eugene Parker 
are also gratefully acknowledged.

\end{document}